\documentclass{emulateapj}
\usepackage{apjfonts}
\usepackage{amsmath}
\begin{document}

\title{Neutrinos from gamma-ray bursts:  propagation of cosmic rays in their host galaxies}
\author{Zi-Yi Wang \altaffilmark{1,2}, Xiang-Yu Wang\altaffilmark{1,2}, Jun-Feng Wang\altaffilmark{3}}

\begin{abstract}
Gamma-ray bursts (GRBs) are proposed as candidate sources of
ultra-high energy cosmic rays (UHECRs). We study the possibility
that the PeV neutrinos recently observed by IceCube are produced
by GRB cosmic rays interacting with the interstellar gas in the
host galaxies. By studying the relation between the X-ray
absorption column density $N_{\rm H}$ and the surface
star-formation rate of GRB host galaxies, we find that $N_{\rm H}$
is a good indicator of the surface gas density of the host
galaxies. Then we are able to calculate the neutrino production
efficiency of CRs for GRBs with known $N_{\rm H}$. We collect a
sample of GRBs that have both measurements of $N_{\rm H}$ and
accurate gamma-ray fluence, and attempt to calculate the
accumulated neutrino flux based on the current knowledge about
GRBs and their host galaxies. When the CR intensity produced by
GRBs is normalized with the observed UHECR flux above
$\sim10^{19}{\rm eV}$,  the accumulated neutrino flux at PeV
energies is estimated to be about $(0.3\pm0.2)\times10^{-8}
\rm{GeV\ cm^{-2}\ s^{-1}\ sr^{-1}} $ (per flavor) under the
assumption that GRB energy production rate follows the cosmic
star-formation rate and  the favorable assumption about the CR
diffusion coefficient. {This flux is insufficient to account for
the IceCube observations, but the estimate suffers from some
assumptions in the calculation and thus we can not rule out this
scenario at present.}
\end{abstract}


\affil{$^1$ School of Astronomy and Space Science, Nanjing University, Nanjing, 210093, China;  xywang@nju.edu.cn \\
$^2$ Key laboratory of Modern Astronomy and Astrophysics (Nanjing
University), Ministry of Education, Nanjing 210093, China \\
$^3$Department of Astronomy and Institute of Theoretical Physics
and Astrophysics, Xiamen University, Xiamen, Fujian 361005, China}

\section{Introduction}
Gamma-ray bursts (GRBs) have been proposed as a potential origin
of ultra-high-energy cosmic rays (UHECRs) (e.g., Waxman 1995;
Vietri 1995; Milgrom and Usov 1995). High-energy neutrinos  are
thought to be a useful messenger to probe CR acceleration in GRBs,
as  they are predicted to be produced in the dissipative
fireballs, where cosmic-ray protons interact with fireball
gamma-ray photons through photopion process (e.g., Waxman \&
Bahcall 1997; Guetta et al. 2004; Dermer \& Atoyan 2006;  Murase
et al. 2006; Wang\& Dai 2009). However, search for high-energy
neutrinos in coincident with GRBs using IceCube has failed to find
any associated neutrinos so far (Abbasi et al. 2012; Aartsen et
al. 2015). The non-detection has put stringent constraints on the
neutrino production efficiency and fireball properties of GRBs (He
et al. 2012; Zhang \& Kumar 2013;  Gao et al. 2013; Yacobi et al.
2014). Because the low neutrino flux could be simply due to a low
efficiency in converting CRs to neutrinos, as expected in some GRB
models with large dissipation radius,  one can not rule out CR
acceleration in GRBs with the non-detection of neutrinos.

CRs accelerated by GRBs will finally escape from the source and
enter the interstellar space of the host galaxy. The proton-proton
collisions between CRs and nuclei in the interstellar medium will
also produce high-energy neutrinos. Recently, IceCube
Collaboration  reported the detection of extraterrestrial TeV-PeV
neutrinos with a best-fit  flux of
$E_\nu^2\Phi_\nu=(0.95\pm0.3)\times10^{-8} \rm{GeV\ cm^{-2}\
s^{-1}\ sr^{-1}} $ per flavor (Aartsen et al. 2014). Noting the
coincidence between the IceCube neutrino flux and the
Waxman-Bahcall bound ($\sim10^{-8} \rm{GeV cm^{-2} s^{-1} sr^{-1}}
$) derived from the flux of UHECRs above $10^{19}{\rm eV}$, some
authors suggest that IceCube neutrinos may be produced by the same
source responsible these UHECRs (Waxman 2013; Katz et al. 2013).
As GRBs are one candidate source of UHECRs, one may wonder whether
neutrinos resulted from the CR collisions with the ISM in the host
galaxy can explain the IceCube observations (Waxman 2013, Wang et
al. 2014). In this paper, we study this interesting possibility by
perform a calculation of the expected neutrino flux with the
knowledge about the properties of GRBs and their host galaxies.

One key unknown factor that decides the neutrino flux from GRB
host galaxies is the pion production efficiency, i.e. the energy
loss of CR protons into pions due to collisions with the ISM of
the host galaxy. CRs interact with the ISM along their path before
escape, so this efficiency relates with the gas column density
that CRs traversed.  As CRs are transported outward by galactic
winds, the gas surface density $\Sigma_{\rm g}$ represents an
averaged column density of matter that CRs traversed. However, for
most GRBs, we do not have measurements of $\Sigma_{\rm g}$ at the
GRB explosion site as GRB hosts are hardly spatially resolved. We
note that the X-ray absorption column density $N_{\rm H}$,
inferred from the X-ray observations of the prompt and afterglow
emission, has a similar role, but $N_{\rm H}$ reflects the column
density along the direction of the line of sight (rather than the
averaged gas column density). Moreover, the value of $N_{\rm H}$
is derived assuming solar metallicity for the absorbing gas, while
GRBs seem to occur preferentially in low metallicity galaxies
(Stanek et al. 2006; Prieto et al. 2008). The X-ray absorption, as
measured with Swift/XRT observations, mainly takes place through
inner shell electrons of metals, thus it is linked to the
metallicity of the ISM in the host galaxies. If GRB host galaxies
have lower metallicity, the true absorbing gas column density
should be higher than the quoted values obtained from fitting
X-ray afterglows, considering this metallicity difference.  We
thus first study how well $N_{\rm H}$ can trace the gas surface
density $\Sigma_{\rm g}$ in \S 2. We find that $N_{\rm H}$ is well
correlated with the surface star-formation rate ($\Sigma_{\rm
SFR}$) of the host galaxies. Then taking into account the
Kennicutt - Schmidt law that relates the surface gas density and
the surface star-formation rate (Kennicutt 1998), we obtain a
relation between $N_{\rm H}$ and $\Sigma_{\rm g}$. Once
$\Sigma_{\rm g}$ is known for each GRB, we can calculate the
neutrino production efficiency and further calculate the
accumulated neutrino flux from all GRB host galaxies (\S 3).

\section{The relations between $N_{\rm H}$ and  $\Sigma_{\rm g}$}
Because we have very few measurements of $\Sigma_{\rm SFR}$, but
have measurement of $N_{\rm H}$ for many GRBs,  we study whether
$N_{\rm H}$ are correlated with $\Sigma_{\rm SFR}$ and act as an
indicator of the gas surface density. We collect the GRB hosts
that have spatially resolved observations from literatures
(Svensson et al. 2010; Fruchter et al. 2006; Savaglio et al.
2009), which are given in Table 1. The total SFR and $80\%$ light
radius of these hosts $R_{80}$ are obtained in these observations.
By comparing the $80\%$ light radius and half light radius of some
GRB hosts  that are spatially resolved (Fruchter et al. 2006;
Kelly et al. 2014), we find that $R_{80}\simeq 2R_{50}$.
Converting $80\%$ light radius to the half-light radius $R_{50}$
with $R_{80}= 2R_{50}$, we calculate the surface star-formation
rates $\Sigma_{\rm SFR}$. With a sample of 18 GRBs that have both
measurements of $N_{\rm H}$ from the XRT data of Swift
observations and $\Sigma_{\rm SFR}$ of their host galaxies, we
then study the relation between $N_{\rm H}$ and
$\Sigma_{\rm{SFR}}$. Through the ordinary least-squares (OLS)
bisector fitting\footnote{Since our goal is to estimate the
functional relation between $N_{\rm H}$ and $\Sigma_{\rm{SFR}}$,
and it is not clear which variable is causative. Therefore, the
OLS bisector fitting which treat  the variables symmetrically
should be used, according to Isobe et al. (1990).} (Isobe et al.
1990), we find that
\begin{equation}
\Sigma_{\rm{SFR}}=10^{-1.89\pm0.27}\left( \frac {N_{\rm
H}}{10^{21} \rm{cm}^{-2}}\right)^{1.42\pm0.25}M_\odot\rm{
yr}^{-1}\rm{ kpc}^{-2}
\end{equation}
with a correlation coefficient of $r=0.756$ and a null hypothesis
probability of $p=7.09\times10^{-4}$, indicating a good
correlation. The result of the fit is shown in the left panel of
Fig.1. Considering the Kennicutt-Schmidt law between the surface
SFR and the surface density of  molecular plus atomic gas
(Kennicutt 1998), i.e.,
\begin{equation}
{\Sigma _{{\rm{SFR}}}} = \left( {2.5 \pm 0.7} \right) \times {10^{
- 4}}{\left( {\frac{{{\Sigma _g}}}{{1{M_ \odot }{\rm{
p}}{{\rm{c}}^{ - 2}}}}} \right)^{1.4 \pm 0.15}}{M_ \odot }{\rm{
y}}{{\rm{r}}^{ - 1}}{\rm{ kp}}{{\rm{c}}^{ - 2}},
\end{equation}
we get
\begin{equation}
{\Sigma_{\rm
g}}=\left(2.1\pm1.0\right)\times{10^{21}\left(\frac{N_{H}}{10^{21}\rm{cm}^{-2}}\right)^{1.01\pm0.21}m_{\rm
H}\rm{cm}^{-2}},
\end{equation}
where $m_{\rm H}$ is the mass of the hydrogen atom. The roughly
linear relation between ${\Sigma_{\rm g}}$ and $N_{\rm H}$
suggests that $N_{\rm H}$ is a good indicator of the gas surface
density $\Sigma_{\rm g}$. Note that $N_{\rm H}$ is derived
assuming solar metallicity for the absorbing gas. Since GRB host
galaxies usually have  lower metallicity, the true absorbing gas
column density should be corrected by the metallicity effect. The
coefficient $2.1\pm1.0$ may reflect such a correction.

We also study whether  $N_{\rm H}$ are correlated with the total
SFRs of the host galaxies.  We collect 35 GRBs in total that have
measurements of $N_{\rm H}$ and total SFRs, which are also listed
in Table 1. Through the OLS bisector fitting of the data of
$N_{\rm H}$ and SFR, we find that
\begin{equation}
 {\rm{SFR}} = {10^{ - 0.63\pm 0.23}}{\left(
{\frac{{{N_{\rm H}}}}{{{{10}^{21}}{\rm{c}}{{\rm{m}}^{ - 2}}}}}
\right)^{1.55 \pm 0.26}}{M_ \odot }{\rm{ yr}}^{ - 1},
\end{equation}
with a correlation coefficient of $r=0.5576$ and a null hypothesis
probability $p=0.0014$, which indicates a tight positive
correlation. The result of the fitting is shown in the right panel
of Fig.1. With this relation, we can infer the SFR of GRB hosts
with $N_{\rm H}$ for each GRB in our sample, which will be used to
compute the galactic wind velocity in \S 3.1.

\section{Neutrino flux from GRB host galaxies}
\subsection{The pion production efficiency}
The GRB accelerated CR protons travel through the host galaxy and
produce high-energy  neutrinos via
proton--proton(\emph{pp})-collisions with its interstellar medium
(ISM). The collisions produce charged pions, which decay to
neutrinos ($\pi ^{+}\rightarrow \nu _{\mu }\bar{\nu}_{\mu }\nu
_{e}e^{+},\,\pi ^{-}\rightarrow \bar{\nu}_{\mu }\nu _{\mu
}\bar{\nu}_{e}e^{-}$). Meanwhile, CRs can escape out of the galaxy
through diffusion or galactic wind advection. These two competing
processes (i.e. collision and escape) regulate the efficiency of
the pion-production of CRs, which can be described by $f_{\pi
}=1-\exp \left( -t_{\text{esc}}/t_{\text{ loss}}\right) $, where
$t_{ \text{loss}}$ is the energy-loss time of CRs via ($pp$)
collisions and $t_{\text{esc}}$ is the escape time of CRs. The
\emph{pp}-collision energy loss time is
$t_{\rm{loss}}=\left(\kappa n\sigma_{pp}c\right)^{-1}=\left(\kappa
{\Sigma_{\rm g}}\sigma_{pp}c/{l}\right)^{-1}$, where $\kappa\simeq
0.5$ is the inelasticity, $n$ is the gas number density, $l$ is
the scale height  of the gaseous disc of the galaxy.
${{\sigma_{{\rm{\emph{pp}}}}}}$ is the \emph{pp}-collision
inelastic cross-section, which slightly increases with the proton
energy, given by
$\sigma_{pp}=\left(34.3+1.88L+0.25L^2\right)\times\left[1-\left(\frac{1.22\times10^{-6}}{E_p/1\rm{GeV}}\right)^4\right]^2{\rm
mb}$ where $L={\rm{log}}\left(\frac{E_p}{1\rm{TeV}}\right)$
(Kelner et al. 2006). Thus, the collision energy loss time is
$t_{\rm{loss}}=5.4\times10^6{\rm{yr}}\frac{l}{500\rm{pc}}
\left(\frac{\Sigma_{\rm g}}{0.01 {\rm g
cm^{-2}}}\right)^{-1}\left(\frac{\sigma_{pp}}{100\rm{mb}}\right)^{-1}$.

The CRs escape from the GRB host galaxy in two ways. One is
advective transportation by the galactic wind.  Galactic-scale
gaseous winds are ubiquitous in star-forming galaxies at all
cosmic epochs (Heckman et al. 1990; Pettini et al. 2001; Shapley
et al. 2003). Such winds can be driven by stellar winds, supernova
explosions or other processes. It is found that,  for luminous and
ultraluminous infrared galaxies at low redshifts, the winds from
more luminous starbursts have higher speeds, roughly as
$v_{w}\propto $ SFR$^{k} $ with $k$ being in the range of
0.25-0.35 (Martin 2005; Rupke et al. 2005; Arribas et al. 2014). A
similar trend is found for star-forming galaxies at $z\sim1$
(Weiner et al. 2009), i.e.
\begin{equation}
v_{w}\approx 175\left( \frac{\text{SFR}}{1 M_{\odot }\text{
yr}^{-1}}\right) ^{0.3},
\end{equation}
with the error in $v_w$ being $35\%$. Such type of relation
between the velocity of the outflowing material and SFR is
expected if the mechanical energy is supplied by stellar winds and
supernova explosions. Thus, one can calculate $v_w$ for each GRB
hosts with measured $N_{\rm H}$ under the help of Eq. 4 and Eq. 5.
Taking $v_w=500\ {\rm{km\ s}}^{-1}$ as the reference value for GRB
host galaxies, the advective escape time is $ t_{\rm{adv}}= l/v_w=
9 \times {10^5}{\rm{yr}}\frac{l}{{500{\rm{pc}}}} {\left(
{\frac{{{v_w}}}{{{500}{\rm{km}}{{\rm{s}}^{ - 1}}}}} \right)^{ -
1}}$.

The other way that CRs escape is diffusion, i.e., CRs are
scattered off small-scale magnetic field inhomogeneities randomly
and diffuse out of the host galaxy. The diffusion time is
estimated to be $t_{\rm{diff}}= l^2/4D$, where
$D=D_0\left(E/E_0\right)^\delta$ is the diffusion coefficient, and
$D_0$ and $E_0=3\rm{GeV}$ are normalization factors. Since little
is known about the diffusion coefficient in GRB host galaxies, in
the calculation we allow  lower diffusion coefficients than that
of our Galaxy, which is $D_0=10^{28}{\rm{cm}}^2{\rm{s}}^{-1}$. The
energy dependence of the diffusion coefficient is also unknown and
$\delta=0-1$ depending on the spectrum of interstellar magnetic
turbulence. We assume two cases, one is the commonly-used value
$\delta=0.5$, based on the measurement of the CR confinement time
in our Galaxy (Engelmann et al. 1990; Webber et al. 2003). Another
choice is $\delta=0.3$, assuming the Kolmogorov-type turbulence.
Then the diffusive escape time is ${t_{{\rm{diff}}}}
={10^4}{\rm{yr}}{\left( {\frac{l}{{500{\rm{pc}}}}} \right)^2}
{\left(
{\frac{{{D_0}}}{{{{10}^{28}}{\rm{c}}{{\rm{m}}^2}{{\rm{s}}^{ -
1}}}}} \right)^{ - 1}}{\left( {\frac{{{\varepsilon
_p}}}{{60{\rm{PeV}}}}} \right)^{ - 0.3}}$, with
$\varepsilon_p\simeq25(1+z)\varepsilon_\nu$.

Combining the advective and diffusive escape timescales, the total
escape time is
$t_{\rm{esc}}^{-1}=t_{\rm{diff}}^{-1}+t_{\rm{adv}}^{-1}.$  When
the difference between $t_{\rm{loss}}$ and $t_{\rm{adv}}$ is
large,
$t_{\rm{esc}}\thickapprox{\rm{min}}\left(t_{\rm{diff}},t_{\rm{adv}}\right)$.
Note that when $t_{\rm{adv}}\ll t_{\rm{diff}}$, which holds for
low-energy CRs and small values of $D_0$, the pion production
efficiency is independent of $l$ and is given by
\begin{equation}
f_\pi\simeq0.17\left(\frac{\Sigma_{\rm g}}{0.01{\rm g
cm^{-2}}}\right) \left(\frac{v_w}{500{\rm km
s^{-1}}}\right)^{-1}\left(\frac{\sigma_{pp}}{100\rm{mb}}\right).
\end{equation}
The dependence of $f_\pi$ on  CR energy has a break at the energy
where $t_{\rm{adv}}=t_{\rm{diff}}$. Below the break energy,
$f_{\pi}$ slightly increases with the energy due to the
\emph{pp}-collision inelastic cross-section, while above the break
$f_{\pi}$ decreases as $\varepsilon_p^{-\delta}$, leading to a
steeper neutrino spectrum.

\subsection{Calibration of CR flux from GRBs}

The Fermi/GBM detection rate of GRBs is about 250 per year
(Paciesas et al. 2012), and the total gamma-ray (10-1000 keV)
fluence   of GRBs detected in one year is $2\times10^{-3}{\rm erg
cm^2}$. The GBM field of view (FOV) is roughly $\Omega_{\rm GBM} =
2\pi$ steradians, so the total all-sky flux of gamma-rays in
10-1000 KeV is $\digamma_\gamma=4\times10^{-3}{\rm erg cm^{-2}
yr^{-1}}$. Then the energy production rate of gamma-rays in the
energy range of 10-1000 keV, is estimated to be $W_\gamma(z=0)=\xi
\digamma_\gamma H/c=5\times10^{42}(\xi/0.5){\rm erg Mpc^{-3}
yr^{-1}}$, where $H$ is the Hubble constant and $\xi$ is a factor
accounting for the source density evolution with redshift (Eichler
et al. 2010). According to the estimate of Katz et al. (2009), the
present-day {\em differential} energy production rate of UHECRs
above 30 EeV is $\varepsilon_p^2
d\dot{n}/d\varepsilon_p(z=0)=(0.45\pm0.15)(\alpha-1)\times10^{44}
{\rm erg Mpc^{-3} yr^{-1}}$, where $\alpha$ is the power-law index
of the CR spectrum. We assume $\alpha=2$ as suggested by Fermi
acceleration.  Define $\eta_p$ as the ratio between the {\em
differential} energy production rate of UHECRs and the energy
production rate of gamma-rays in 10-1000 keV. If GRBs are the
source of UHECRs above 30 EeV, $\eta_p=(\varepsilon_p^2
d\dot{n}/d\varepsilon_p)/W_\gamma(10-1000 KeV)=9\pm3$, assuming
$\xi=0.5$ (Eichler et al. 2011).

\subsection{The neutrino flux}
We collect a sample of GRBs that have both measurements of $N_{\rm
H}$ and  gamma-ray fluence. As the energy coverage of Swift BAT is
small, we choose  GRBs that are detected by Fermi/GBM to get more
accurate values of the gamma-ray fluence. There are 45 GRBs in
total that have measurements of both $N_{\rm H}$ and gamma-ray
fluence, as shown in Table 2. For each GRB, the (single-flavor)
neutrino fluence is estimated to be
\begin{equation}
\varepsilon_\nu^2\phi_\nu=\frac{1}{6}{f_\pi }\varepsilon
_{p}^2{\phi_p}=\frac{1}{6}f_\pi\eta_p {F_\gamma},
\end{equation}
where $\varepsilon _{p}^2{\phi_p}$ is the differential energy flux
of CR protons  and $F_\gamma$ is the gamma-ray fluence of this
GRB. The accumulated neutrino flux produced by all GRBs in the
universe is
\begin{equation}
\varepsilon_\nu^2\Phi_\nu=\frac{1}{N}\sum\limits_{i}\varepsilon_\nu^2\phi_{\nu,i}R_{\rm
GRB}{\left( {4\pi } \right)^{ - 1}}
\end{equation}
where $i$ represents the $i$th GRB in our sample,  $N$ is the
total number of GRBs in our sample, and $R_{\rm GRB}=500{\rm
{yr}}^{-1}$ is the all-sky  rate of  GRBs that would be detected
by Fermi/GBM if the field of view of GBM is $4\pi$ steradians. The
result of the accumulated neutrino flux is shown in Fig.2. We find
that the accumulated neutrino flux at PeV is
$(0.3\pm0.2)\times10^{-8} \rm{GeV\ cm^{-2}\ yr^{-1}\ sr^{-1}} $
per flavor for $\delta=0.3$ and $D_0=10^{26} {\rm cm^2 s^{-1}}$.
The 1 sigma error in the accumulated flux  is obtained by using
formulas for the propagation of error. It results mainly  from the
uncertainties in the neutrino production efficiency $f_\pi$, in
the values of $N_{\rm H}$, and in the CR-to-gamma-ray ratio
$\eta_p$.

Since the CR diffusion coefficient is not well-understood, we
study its effect on the neutrino flux. The results for different
values of $\delta$ and $D_0$ are shown in Fig. 3. One can see
that, when $D_0$ decreases, the neutrino flux increases at PeV
energies. For $\delta=0.3$, if $D_0$ is as low as $10^{25}{\rm
cm^2 s^{-1}}$, the neutrino spectrum becomes too hard, although
the flux increases. For $\delta=0.5$, lower values of $D_0$ is
needed to produce the same flux compared to the case of
$\delta=0.3$. However, the confinement of 100 PeV CRs producing
PeV neutrinos requires that $D_0$  be larger than $10^{24 } {\rm
cm^2 s^{-1}}$ (Murase et al. 2013). Thus, even under favorable
conditions about the CR diffusion coefficient, the neutrino flux
produced by GRB CRs alone can not explain the IceCube observations
{if other assumptions used in our calculation are correct.}

In the  calculation of the accumulated neutrino flux, we only
considered those GRBs that triggered Fermi/GBM detector (i.e.
adopting $R_{\rm GRB}=500 {\rm yr^{-1}}$). There are dim GRBs that
do not trigger the detector and their total gamma-ray fluence  may
even be larger than that of the triggered ones according to the
simulation results in Liu \& Wang (2013). These untriggered GRBs
may also produce diffuse neutrinos as triggered GRBs. However, one
should note that considering the contribution by  untriggered GRBs
would not change our result about the accumulated neutrino flux,
because the the CR-to-gamma-ray ratio $\eta_p$ needs to be
re-calibrated correspondingly and thus the accumulated neutrino
flux remains unchanged.

There could be an exceptional case, i.e., if some dim GRBs do not
accelerate protons to energy $10^{19}$ eV (although nuclei can
still be accelerated to ultra high energies), but they can still
contribute to PeV neutrinos with CRs of $\ga100{\rm PeV}$. {Note
also that this does not exclude the possibility that
low-luminosity GRBs themselves are responsible for their origin
(e.g., Murase \& Ioka 2013, Bhattacharya et al. 2014).}

\section{Discussions and Conclusions}

{The above calculation has some uncertainties in the following
aspects. First, the estimate of the energy production rate of
gamma-rays $W_\gamma$ has uncertainty (e.g., Dermer 2012). Note
that the factor $\xi=0.5$   is obtained by assuming that GRB rate
follows the cosmic SFR of Porciani \& Madau (2001) or that of
Hopkins \& Beacom (2008). If GRB density evolves faster than the
cosmic SFR, $\xi$ is smaller and then $\eta_p$ is larger. Second,
the estimate of the UHECR energy budget  has uncertainty. The
value can be a bit larger if one uses the Telescope Array or HiRes
data. Third, the power-law index $\alpha$ could be softer than 2,
then the  energy budget of CRs at 100 PeV could be larger. Thus,
given these uncertainties, the neutrino flux in the optimistic
case could reach the observed value of IceCube.}

{On the other hand,   if one consider that GRB host galaxies that
are not detected by optical observations are possibly smaller
galaxies, such as dwarf galaxies, the pion production efficiency
may be smaller and thus the neutrino flux contributed by these
galaxies would be smaller.} Also, as shown by Eq. 6, the pion
production efficiency depends on the speeds of the galactic winds.
The properties of the galactic winds in GRB host galaxies are not
well-explored. X-ray observations of starburst galactic winds,
such as M82, usually give a higher speeds than that inferred from
the optical observations (Strickland \& Heckman 2009). If the
galactic wind speeds of GRB hosts are proven to be higher, the
neutrino flux produced by GRB hosts would decrease. {The pion
production efficiency also depends on the diffusive coefficient.
If $D_0 $ is larger than  $10^{27}\rm{cm}^2\rm{s}^{-1}$ (for
example, $D_0=10^{28}\rm{cm}^2\rm{s}^{-1}$ for our Galaxy), the
accumulated neutrino flex would be lower than $10^{-9} \rm{GeV\
cm^{-2}\ yr^{-1}\ sr^{-1}} $ at $1\rm{PeV}$, as shown in Fig. 3.}

In summary, we calculated the neutrino flux produced by CRs
accelerated by GRBs while they are propagating in the host
galaxies based on our current knowledge about GRB and their host
galaxies. These CRs collide with nuclei of ISM and produce
neutrinos before they escape out of the galaxy. When the flux of
CRs produced by GRBs is normalized with the observed flux of
UHECRs above $\sim10^{19}{\rm eV}$, the accumulated neutrino flux
is $(0.3\pm0.2)\times10^{-8} \rm{GeV\ cm^{-2}\ s^{-1}\ sr^{-1}} $
per flavor under the usual assumptions about the GRB properties
and favorable assumptions about the CR diffusion coefficient. {The
estimate, however, has uncertainty due to uncertainty in our
current knowledge of GRB and their host galaxies  and the
accumulated neutrino flux could reach the observed value by
IceCube in the optimistic case, so we can not  rule out this
scenario at present.}

\acknowledgments We thank Zhuo Li and Ruoyu Liu for useful
discussions. This work is supported by the 973 program under grant
2014CB845800, the NSFC under grants 11273016 and 11033002, and the
Excellent Youth Foundation of Jiangsu Province (BK2012011).

\clearpage

\begin{figure}
\plottwo{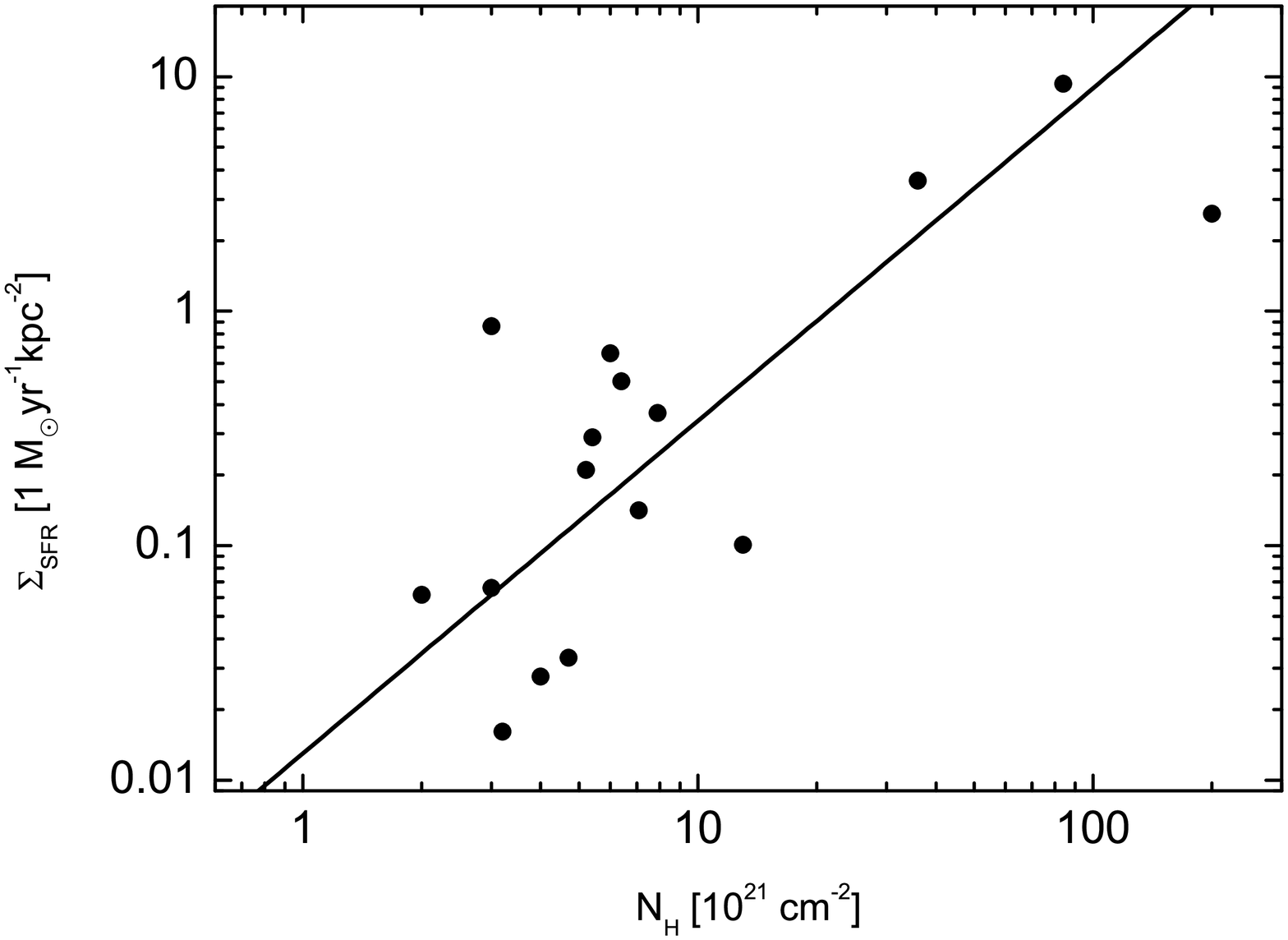}{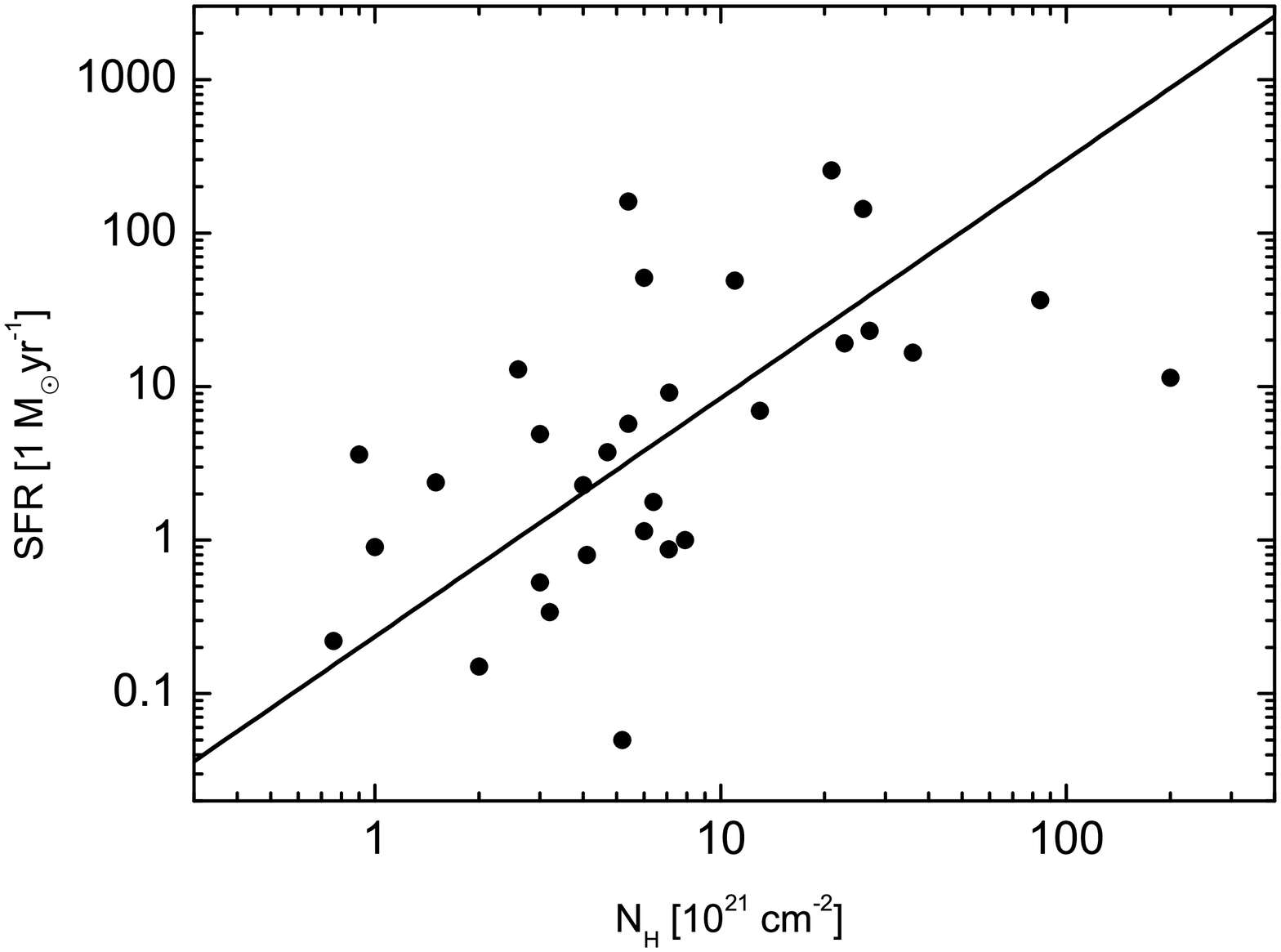}\\
  \caption{Left panel: the relation between the surface star-formation rate $\Sigma_{\rm{SFR}}$ and the X-ray absorption column density $N_{\rm H}$.
  Right panel:  the relation between the total star-formation rate ${\rm{SFR}}$ and the X-ray absorption column density $N_{\rm H}$.
  The data  points are from table 1.   The lines represent the best OLS bisector fits. See the text for more details. }\label{1}
\end{figure}

\begin{figure}
\epsscale{1} \plotone{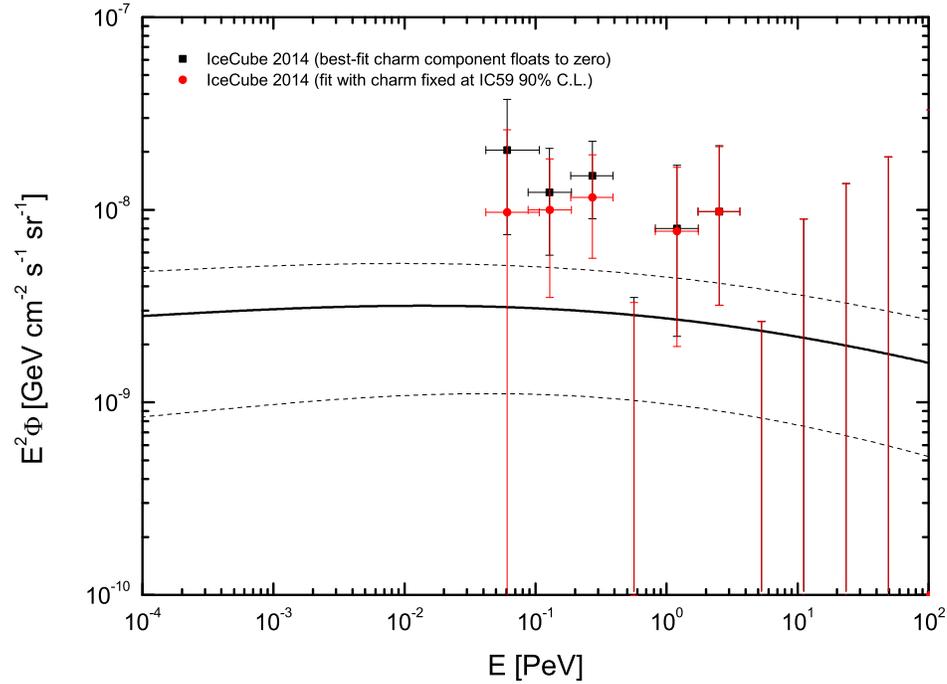}\\
  \caption{ The energy spectra of neutrino emission in comparison with IceCube data. The solid line represents the neutrino flux obtained using $\delta=0.3$ and $D_0=10^{26}\rm{cm^2\ s^{-1}}$.
and  the dashed lines are the $1-\sigma$ error. The IceCube data
are also shown. }\label{2}
\end{figure}

\begin{figure}
\epsscale{1} \plotone{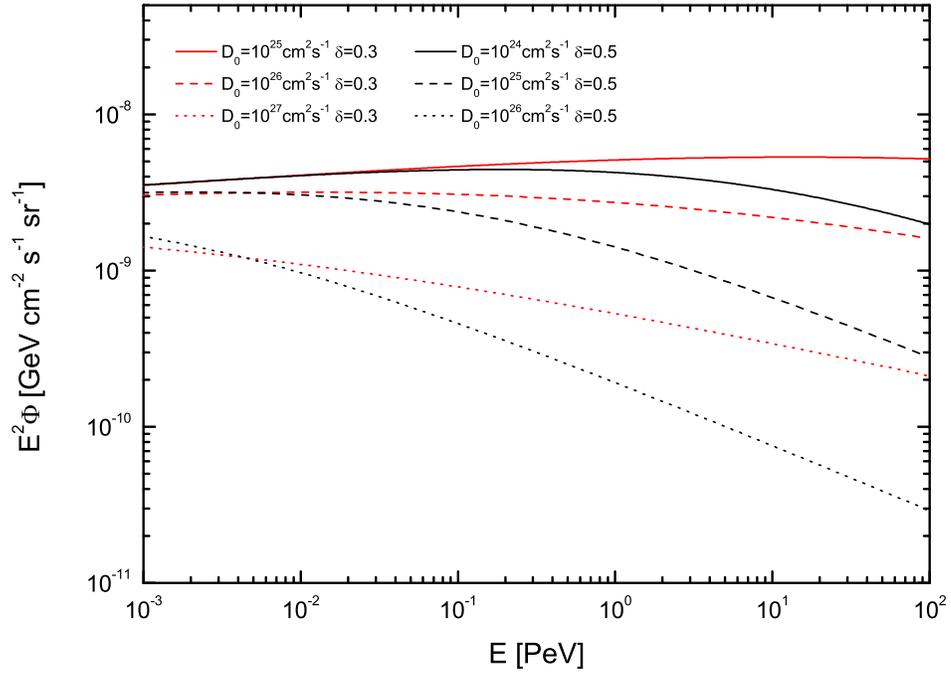}\\
  \caption{The energy spectra of neutrino emission for different values of $\delta$ and $D_0$. }\label{3}
\end{figure}

\begin{table}
\centering \caption{Galaxy parameters of GRB hosts }
\begin{tabular}{lllll}
\hline
GRB &    SFR [$\rm M_\odot\ yr^{-1}$]   &   $\rm N_{\rm H}$ [$10^{21}\rm cm^{-2}$]  &   $\rm R_{80}$ [kpc]   & Ref. \\
\hline
GRB 050826  &   9.13 &   $7.1^{+2.0}_{-2.0}$ & - &[1,2]\\
GRB 051006  &   51  &   $6^{+3}_{-3}$ &- &[3,A]\\
GRB 060814  &   256 &   $20.9^{+2.3}_{-2.2}$ &  -&[3,4] \\
GRB 061121  &   160 &   $5.4^{+0.8}_{-0.5}$ &-&[3,4]\\
GRB 061126  &   2.38&   $1.5^{+0.2}_{-0.2}$ &-&[1,A]\\
GRB 070306  &   143 &  $26.8^{+4.7}_{-4.3}$ & -&[3,4]\\
GRB 080325  &   12.9&   $2.6^{+0.1}_{-0.1}$ &-&[5,A]\\
GRB 080607  &   19.1 &   $22.8^{+4.7}_{-5.2}$ &-& [5,4]\\
GRB 081109  &   49.0 &   $11^{+1.3}_{-1.2}$ &  -&[5,6] \\
GRB 090328  &   4.8 &   $0.9^{+1.1}_{-0.9}$ & -&[7,A] \\
GRB 090424  &   0.8 &   $4.1^{+0.6}_{-0.5}$ &  -&[8,4] \\
GRB 091127  &   0.22 &   $0.76^{+0.35}_{-0.5}$ &-&[9,4]\\
GRB 120624B &   23  &   $27^{+46}_{-26}$ & -& [10,11]\\
GRB 130427  &   0.9 &   $1.0^{+0.1}_{-0.1}$ &-&[10,12]\\
GRB 970228  &   0.53 &   $3^{+6}_{-4}$ & 3.2 &[1,13,14]\\
GRB 970508  &   1.14  &   $6^{+10}_{-5}$  & 1.48 &[1,13,14] \\
GRB 970828  &   0.87  &   $7.08^{+3.21}_{-2.75}$ & 2.8 &[1,15,14]\\
GRB 971214  &   11.4  &   $200^{+290}_{-120}$ & 2.36&[1,13,14]\\
GRB 980703  &   16.57 &   $36^{+22}_{-13}$  & 2.42 &[1,13,14] \\
GRB 990123  &   5.72  &   $5.4^{+9.4}_{-2.7}$ & 5.01 &[1,13,14]\\
GRB 990705  &   6.96  &   13 &  9.38 &[1,16,14] \\
GRB 000926  &   2.28  &   $4^{+3.5}_{-2.5}$ & 10.25 &[1,17,14] \\
GRB 011211  &   4.90  &   $3.0^{+1.1}_{-1.1}$  &  2.69 &[1,18,14] \\
GRB 020405  &   3.74  &   $4.7^{+3.7}_{-3.7}$ & 11.96 &[1,19,14] \\
GRB 030323  &   1     &   $7.9^{+1.4}_{-1.2}$ & 1.86&[20,20,14] \\
GRB 041006  &   0.34  &   $3.2^{+0.16}_{-0.16}$ & 5.19 &[1,17,14]\\
GRB 050416A &   1.77  &   $6.4^{+0.9}_{-0.5}$ & 2.12 &[21,4,21] \\
GRB 050525A &   0.15  &   $2.0^{+0.9}_{-0.9}$ & 1.76  &[21,4,21]\\
GRB 051022  &   36.46 &   $83.9^{+8.5}_{-7.8}$ &  2.23  &[1,5,21] \\
GRB 060218  &   0.05  &   $5.2^{+0.5}_{-0.5}$ & 0.55 &[1,2,21]\\
\hline

\end{tabular}

\
(1)Savaglio, S., et al.\ 2009, \apj, 691, 182 (2)Campana, S., et al.\ 2010, \mnras, 402, 2429 (3)Perley, D.~A.,  et al.\ 2014, arXiv:1407.4456 (4)Campana, S.,  et al.\ 2012, \mnras, 421, 1697 (5)Perley, D.~A.,  et al.\ 2013, \apj, 778, 128 (6)Kr{\"u}hler, T., et al.\ 2011, \aap, 534, AA108 (7)McBreen, S.,  et al.\ 2010, \aap, 516, AA71 (8)Jin, Z.-P.,  et al.\ 2013, \apj, 774, 114 (9)Vergani, S.~D.,  et al.\ 2011, \aap, 535, AA127 (10)Wang, F.~Y., \& Dai, Z.~G.\ 2014, \apjs, 213, 15 (11)de Ugarte Postigo, A.,  et al.\ 2013, \aap, 557, LL18 (12)Littlejohns, O.~M.,  et al.\ 2014, arXiv:1412.6530 (13)Galama, T.~J., \& Wijers, R.~A.~M.~J.\ 2001, \apjl, 549, L209 (14)Svensson, K.~M., \ 2010, \mnras, 405, 57 (15) Yoshida, A.,  et al.\ 2001, \apjl, 557, L27 (16)Amati, L., et al.\ 2000, Science, 290, 953 (17)Kann, D.~A.,  et al.\ 2006, \apj, 641, 993 (18)Reeves, J.~N.,  et al.\ 2003, \aap, 403, 463 (19)Covino, S.,  et al.\ 2003, \aap, 400, L9 (20)Vreeswijk, P.~M.,  et al.\ 2004, \aap, 419, 927 (21)Fruchter, A.~S.,  et al.\ 2006, \nat, 441, 463
(A) is from the GCN website.
\end{table}

\begin{table}
\centering \caption{The X-ray absorption column density and
10-1000KeV fluence}
\begin{tabular}{llllll}
\hline
GRB & $\rm N_{\rm H}$ \tablenotemark{a} [$10^{21}\ \rm cm^{-2}$] &   GBM Fluence\tablenotemark{b} [$\rm erg/cm^{2}$] &z\tablenotemark{a} \\
\hline
141004A &   $1^{+1.5}_{-1}$ &   $1.18\times 10^{-6}$  &0.57  \\
140907A &   $0.7^{+0.7}_{-0.6}$ &   $6.45\times 10^{-6}$  &1.21  \\
140703A &   $5.4^{+8}_{-5.4}$   &   $7.62\times 10^{-6}$ &3.14   \\
140512A &   $0.52^{+0.28}_{-0.26}$  &   $2.93\times 10^{-5}$ &0.725   \\
140506A &   $1.8^{+0.4}_{-0.4}$ &   $6.59\times 10^{-6}$ &0.889   \\
140423A &   $6.6^{+3.2}_{-3}$   &   $1.81\times 10^{-5}$ &3.26   \\
140304A &   $0.38^{+0.34}_{-0.31}$  &   $2.24\times 10^{-6}$  &5.28  \\
140213A &   $1.4^{+1}_{-0.9}$   &   $2.12\times 10^{-5}$ &1.2076   \\
140206A &   $10.1^{+2.4}_{-2.3}$    &   $1.55\times 10^{-5}$ &2.74   \\
131105A &   $1.7^{+1.4}_{-1.2}$ &   $2.38\times 10^{-5}$ &1.686   \\
130612A &   $7.7^{+6.6}_{-5.5}$ &   $6.80\times 10^{-7}$ &2.006   \\
130610A &   $2.6^{+2.8}_{-2.6}$ &   $3.54\times 10^{-6}$ &2.092   \\
130420A &   $0.58^{+0.25}_{-0.24}$  &   $1.16\times 10^{-5}$ &1.297   \\
121211A &   $4.59^{+1.07}_{-0.99}$  &   $6.41\times 10^{-7}$ &1.023   \\
121128A &   $11^{+5}_{-4}$  &   $9.30\times 10^{-6}$ &2.2   \\
120922A &   $0.236^{+0.28}_{-0.27}$ &   $8.21\times 10^{-6}$ &3.1   \\
120907A &   $0.46^{+0.31}_{-0.29}$  &   $8.09\times 10^{-7}$ &0.97   \\
120811C &   $0.67^{+0.34}_{-0.31}$  &   $3.45\times 10^{-6}$ &2.67   \\
120729A &   $0.7^{+0.5}_{-0.4}$ &   $5.08\times 10^{-6}$ &0.8   \\
120712A &   $0.14^{+0.31}_{-0.14}$  &   $4.43\times 10^{-6}$ &4.1   \\
120326A &   $0.39^{+0.23}_{-0.22}$  &   $3.26\times 10^{-6}$ &1.798   \\
120119A &   $1.3^{+0.5}_{-0.4}$ &   $3.87\times 10^{-5}$ &1.73   \\
111228A &   $2.51^{+0.49}_{-0.47}$  &   $1.81\times 10^{-5}$  &0.715  \\
111107A &   $3.5^{+6}_{-3.5}$   &   $9.07\times 10^{-7}$ &2.893   \\
110818A &   $0.52^{+0.27}_{-0.24}$  &   $5.15\times 10^{-6}$ &3.36   \\
110213A &   $0.61^{+0.24}_{-0.23}$  &   $9.37\times 10^{-6}$ &1.46   \\
110128A &   $\sim
0$ &   $1.43\times 10^{-6}$ &2.339   \\
101219B &   $0.5^{+0.7}_{-0.5}$ &   $3.99\times 10^{-6}$ &0.5519   \\
100906A &   $8.1^{+2.5}_{-2.5}$ &   $2.33\times 10^{-5}$ &1.727   \\
100816A &   $2.6^{+1.1}_{-1.1}$ &   $3.65\times 10^{-6}$ &0.8039   \\
100814A &   1.6 &   $1.49\times 10^{-5}$ &1.44   \\
100728B &   $4.3^{+3.1}_{-2.5}$ &   $3.34\times 10^{-6}$  &2.1  \\
100728A &   $2.7^{+0.3}_{-0.3}$ &   $1.28\times 10^{-4}$  &1.567  \\
100704A &   $2.1^{+0.9}_{-0.8}$ &   $1.04\times 10^{-5}$  &3.6  \\
100615A &   $11^{+1.3}_{-1.2}$  &   $8.72\times 10^{-6}$  &1.398  \\
091208B &   $8.3^{+4.3}_{-3.4}$ &   $6.19\times 10^{-6}$  &1.063  \\
091127  &   $0.76^{+0.35}_{-0.5}$   &   $2.07\times 10^{-5}$ &0.49   \\
091020  &   $5.8^{+1.7}_{-1.6}$ &   $8.35\times 10^{-6}$ &1.71   \\
090926B &   $13.9^{+1.6}_{-1.5}$    &   $1.08\times 10^{-5}$ &1.24   \\
090516A &   $22.9^{+4}_{-3.9}$  &   $1.72\times 10^{-5}$ &4.1   \\
090424  &   $4.1^{+0.6}_{-0.5}$ &   $4.63\times 10^{-5}$ &0.544   \\
090423  &   $102^{+49}_{-54}$   &   $8.16\times 10^{-7}$ &8   \\
081221  &   $26.1^{+3.8}_{-3.6}$    &   $3.00\times 10^{-5}$  &2.26  \\
080916A &   $8^{+3.2}_{-1.9}$   &   $7.81\times 10^{-6}$  &0.689  \\
080905B &   $22.6^{+5.3}_{-4.7}$    &   $2.91\times 10^{-6}$  &2.374  \\
\hline
\end{tabular}

(a)From the GCN website.
(b)From Fermi-GBM GRB list of detections. http://www.asdc.asi.it/grbgbm/
\end{table}

\end{document}